\documentclass[11pt]{article}
\usepackage{asp2010, graphicx}

\resetcounters

\begin{document}

\title{Observational Constraints, Stellar Models, and {\it Kepler} Data for 
$\theta$\,Cyg, the Brightest Star Observable by {\it Kepler}}
\author{J.A. Guzik$^1$, G. Houdek$^2$, W.J. Chaplin$^3$, D. Kurtz$^4$, R.L.
Gilliland$^5$, F. Mullally$^{6}$, J.F. Rowe$^{6}$, M.R. Haas$^7$, S.T. 
Bryson$^7$, M.D. Still$^{7,8}$, and T. Boyajian$^{9,10}$
\affil{$^1$XTD-2, MS T-086, Los Alamos National Laboratory, Los Alamos, NM  87545, 
USA}
\affil{$^2$Institute of Astronomy, University of Vienna, A-1180, Vienna, Austria}
\affil{$^3$School of Physics and Astronomy, University of Birmingham, Birmingham B15\,2TT, UK}
\affil{$^4$Jeremiah Horrocks Institute, University of Central Lancashire, Preston 
PR1\,2HE, UK}
\affil{$^5$Space Telescope Science Institute, 3700 San Martin Dr., Baltimore, MD  21218, USA}
\affil{$^6$SETI Institute/NASA Ames Research Center, Moffett Field, CA  94035, USA}
\affil{$^7$NASA Ames Research Center, Bldg. 244, MS-244-30, Moffett Field, CA  94035, USA}
\affil{$^8$Bay Area Environmental Research Institute, 560 Third Street W., Sonoma, CA  95476, USA}}
\affil{$^{9}$Center for High Angular Resolution Astronomy/Department of Physics and Astronomy, Georgia State University, P.O. Box 4106, Atlanta, GA  30302, USA}
\affil{$^{10}$Hubble Fellow}

\begin{abstract}
The $V\,=\,4.48$ F4\,main-sequence star $\theta$\,Cyg is the brightest star observable 
in the\,{\it Kepler} spacecraft field of view.  Short-cadence (58.8\,s) photometric 
data were obtained by {\it Kepler} during 2010 June$-$September. Preliminary 
analysis shows solar-like oscillations in the frequency range $1200-
2500$\,$\mu$Hz.  To interpret these data and to motivate further observations, we 
use observational constraints from the literature to construct stellar evolution 
and pulsation models for this star.  We compare the observed large frequency 
separation of the solar-like oscillations with the model predictions, and discuss 
the prospects for $\gamma$\,Doradus-like g-mode pulsations, given the 
observational constraints.  We discuss the value of angular diameter measurements 
from optical interferometry for constraining stellar properties and the 
implications for asteroseismology.
\end{abstract}

\section{Introduction}

The main mission of the NASA {\it Kepler} spacecraft, launched 2009 March 7, is to 
search for Earth-sized planets around Sun-like stars using high precision CCD 
photometry to detect planetary transits (Borucki et al.\,2010).  As a secondary 
mission objective, {\it Kepler} is surveying and monitoring over 10\,000 stars for 
asteroseismology, i.e., using the intrinsic brightness variations caused by 
pulsations to infer the star's mass, age, and interior structure (Gilliland et al. 
2010).\footnote{See http://kepler.nasa.gov for more information about the {\it Kepler} 
mission.}

The $V\,=\,4.48$ F4V star $\theta$\,Cyg (13\,Cyg, HR\,7469, HD\,185395, KIC\,11918630,  
where KIC = {\it Kepler} Input Catalogue) is the brightest star that falls on 
active pixels in the {\it Kepler} field of view.  Because $\theta$\,Cyg is nearby 
and bright, excellent ground-based data can be combined with extremely high 
signal-to-noise long time series {\it Kepler} photometry to provide constraints 
for asteroseismology.  The position of $\theta$\,Cyg in the HR~Diagram is among 
known $\gamma$\,Dor pulsators, suggesting the possibility that it may exhibit 
high-order gravity mode pulsations, which would probe the stellar interior just 
outside its convective core.  $\theta$\,Cyg is also cool enough to exhibit solar-like
p-mode acoustic oscillations, which would probe both the interior and envelope.  $\theta$\,Cyg's projected rotational velocity $v~\sin i$ is  $7$\,km\,s$^{-1}$ (Erspamer \& North 2003).  If the inclination angle of the rotation axis is not too large, $\theta$\,Cyg's slow rotation should simplify mode identification and pulsation modeling, as spherical approximations 
and low-order perturbation theory for the rotational splitting should be adequate.

$\theta$\,Cyg has been the object of recent adaptive optics observations (Desort 
et al. 2009).  It has a resolved binary M-dwarf companion of 
$\sim$0.35\,M$_{\odot}$ with separation 46\,AU.  Following the orbit for nearly an 
orbital period (unfortunately, $100-200$\,y) will eventually give an accurate 
dynamical mass for $\theta$\,Cyg.  Also, the system shows a 150-d quasiperiod in 
radial velocity, suggesting that possibly one or more planets could accompany the 
stars.

The revised Hipparcos parallax of $54.54 \pm 0.15$\,milliarcseconds (van Leeuwen 2007) gives a 
distance of $18.33\,\pm\,0.05$\,pc, which, combined with an estimate of its apparent 
bolometric luminosity (van Belle et al. 2008), gives $\log L= 0.63\,\pm\,
0.03$\,L$_{\odot}$.  Other parameters derived from multi-color photometry and
high-resolution spectroscopy are $T_{\rm  eff} = 6745\,\pm\,150$\,K, $\log g = 4.2 \pm 
0.2$ (cgs), and $\rm{[M/H]} = -0.04$ (Erspamer \& North 2003; Gray et al.\,2003).  
Nordstr\"om et al.\,(2004) give a mass $1.38\,\pm\,0.05$\,M$_{\odot}$, and age\,1.5  
+0.6/-0.7\,Gyr using Str\"omgren photometry plus the Padova stellar model grid.  

$\theta$\,Cyg has also been the object of optical interferometry observations.  
van Belle et al.\,(2008) used the Palomar Testbed Interferometer to identify 350 stars, including $\theta$\,Cyg, that are suitably pointlike to be used for as calibrators for optical long-baseline interferometric observations.  They then used spectral energy distribution (SED) fitting (not the interferometry measurements) based on 91 photometric observations of $\theta$\,Cyg to estimate its angular diameter to be $0.760\,\pm\,0.021$\,milliarcseconds.  Combining this estimate
with the revised Hipparcos parallax, the interferometric radius is $1.50\,\pm\,0.04$\,R$_{\odot}$.  If one uses only the literature $L$ and $T_{\rm eff}$ and their 
associated error estimates, the derived radius is $1.53\,\pm\,0.13$\,R$_{\odot}$.

Optical interferometry has the potential to constrain the radius of 
$\theta$\,Cyg better than spectroscopy and photometry alone.  At this conference 
we learned of unpublished interferometric measurements by T. Boyajian using the 
CHARA array.  Boyajian measures an angular diameter, $0.861 \pm 
0.015$\,milliarcseconds, somewhat larger than the SED fitting estimate
of van Belle et al., and giving a radius $1.697\,\pm\,0.030$\,R$_{\odot}$ and associated $T_{\rm  
eff} = 6381\,\pm\,65$\,K, considerably different from the literature values.  
Additional observations and analyses are necessary to reconcile and refine the 
observed properties of $\theta$\,Cyg for comparison with those that we hope to 
derive from its pulsation frequencies measured by {\it Kepler}.

\begin{table}
\caption{$\theta$\,Cyg Models and Observational Constraints}
\begin{center}

\begin{tabular}{llll} 
\hline
Property & Observation & Model 1 & Model 2  \\
\hline
Mass (M$_\odot$)& 1.38 $\pm$ 0.05 & 1.38 & 1.29  \\
$\rm{[M/H]}$    & $-$0.04   &       &     \\
$Z^{a}$  &       &   0.017   &   0.013    \\
$T_{\rm eff}$ (K)  & 6745 $\pm$ 150 & 6744  &  6834 \\
$\log L$ (L$_{\odot}$ ) & 0.63 $\pm$ 0.03  &  0.666   &   0.603   \\
Radius (R$_{\odot}$ )     & 1.50 $\pm$ 0.04   &   1.58   &   1.43    \\
$\log g$  & 4.2 $\pm$ 0.2   &   4.18   &   4.24   \\
Core $X^{b}$&      &   0.351   &   0.386   \\
Age (Gyr) &      &   1.61  &   1.63  \\
$T$ CZ$^{c}$ base (K) &      &   494\,800    &    456\,600    \\
\hline
\end{tabular}
\tablenotetext{a}{$Z$ is mass fraction of elements 
heavier than H and He}
\tablenotetext{b}{$X$ is hydrogen mass fraction}
\tablenotetext{c}{CZ = convection zone}
\end{center}
\end{table}

\section{{\it Kepler} $\theta$\,Cyg Observations}

$\theta$\,Cyg was observed 2010 June$-$September ({\it Kepler} Quarter 6).  It is 
the brightest star on active silicon in the {\it Kepler} field of view, and is 7 
magnitudes brighter than the saturation limit.  {\it Kepler} stars are observed 
using one of a set of masks that define the pixels to be stored for that star. For 
$V < 8$, these masks grow prohibitively large.  Special apertures can be defined 
to better conform to the distribution of charge for extremely saturated bright 
stars. For $\theta$\,Cyg the number of recorded pixels  was reduced from  
$>$10\,000 to $\sim$1\,300 by using a special aperture.  Both short-cadence 
(58.8\,s) and long-cadence (29.4\,min) data were obtained.  Fig.\,1 shows the 90-d 
unprocessed light curve.\footnote{The data are available at http://keplergo.arc.nasa.gov/ArchivePublicDataThetaCygni.shtml (see also Haas et al. 2011).}  $\theta$\,Cyg 
was not well-captured by the dedicated mask for $\sim$50\% of the quarter (a 
problem that is now resolved), so 42\,d of the best-quality data were used in the 
pulsation analysis (Fig.\,2).  These data were processed to remove outliers and a small remaining trend before Fourier analysis.   The frequency power spectrum (Fig.\,3) shows a rich spectrum of overtones of solar-like oscillations, with detectable modes in the range from approximately 1200 to 2500\,$\mu$Hz.  The appearance of the oscillation spectrum is very similar to that of two other well-studied F stars:  Procyon~A (Bedding et al. 2010) and HD49933 (Appourchaux et al. 2008).  The envelope of oscillation power is very wide, and modes are evidently heavily damped, meaning the resonant peaks have large widths in the frequency spectrum, which makes mode identification difficult (see, e.g., Bedding \& Kjeldsen 2010).   The frequency of maximum mode amplitude shifts to higher frequency 
with decreasing stellar mass.  For comparison, the peak in the Sun's power 
spectrum is at about 3000\,$\mu$Hz.

It is clear from the {\it Kepler} data that $\theta$\,Cyg is a solar-like 
oscillator.  However, the granulation noise and background are too high at low 
frequencies with the short data set to tell whether there are also modes with low 
frequencies around 11\,$\mu$Hz (or 1\,d$^{-1}$) characteristic of $\gamma$\,Dor
g-mode pulsators (see Grigahc\'ene et al. 2010 and references therein).  It is hoped 
that further observations of $\theta$\,Cyg will reduce the noise level to reveal 
longer-period $\gamma$\,Dor pulsations, as it would be advantageous to find a star 
that shows both $\gamma$\,Dor and solar-like oscillations to help constrain the 
properties of the stellar interior and test the physics of stellar evolution and 
pulsation models.

\begin{figure}%[tb]
\center{\includegraphics[width=0.7\columnwidth]{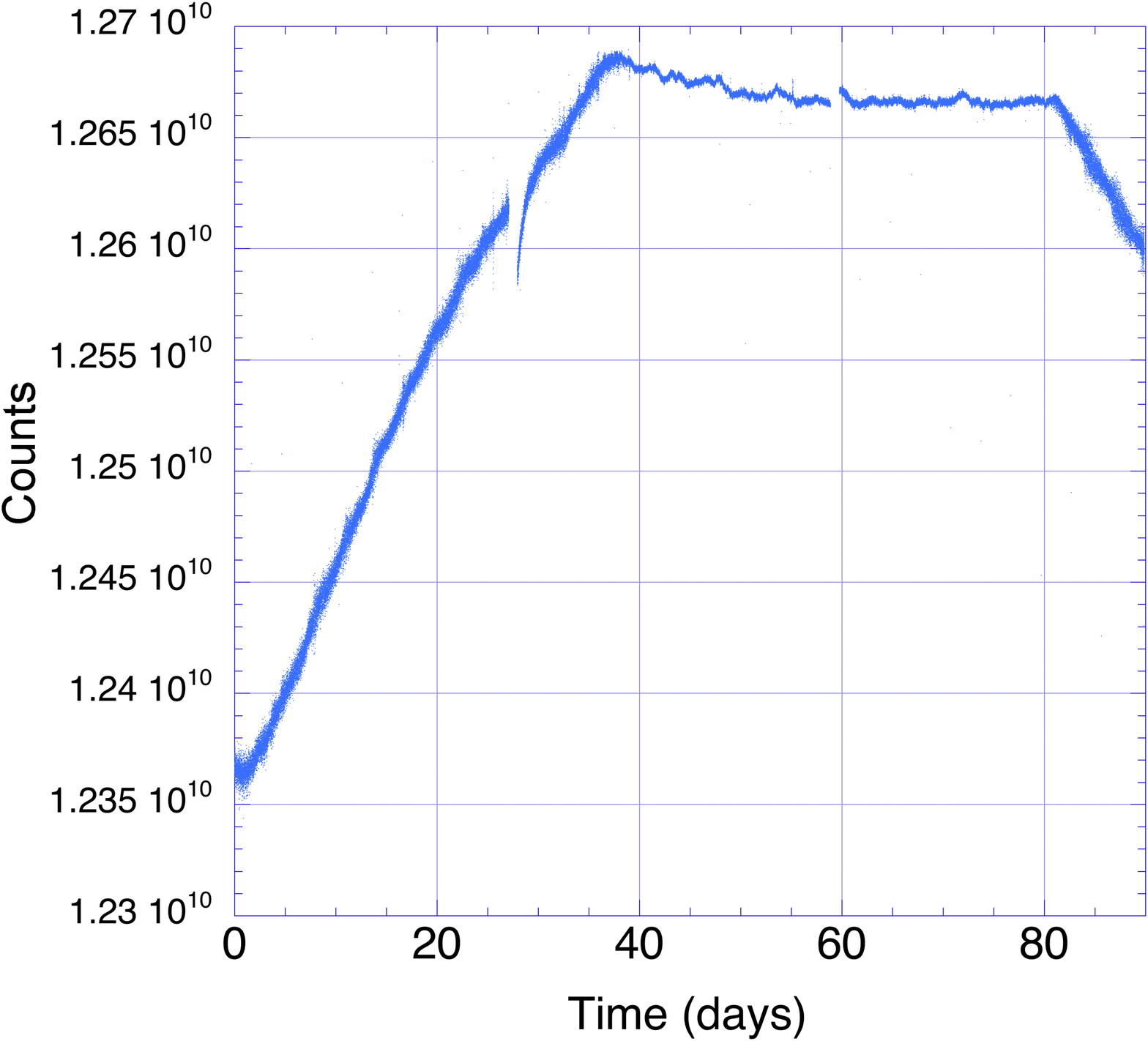}}
\caption{{\it Kepler}  $\theta$\,Cyg light curve for Quarter 6.  The custom 
aperture captured the target completely only during 42 days (approximately days 39 
to 81) used in this analysis} 
\end{figure}

\begin{figure}%[tb]
\center{\includegraphics[width=0.9\columnwidth]{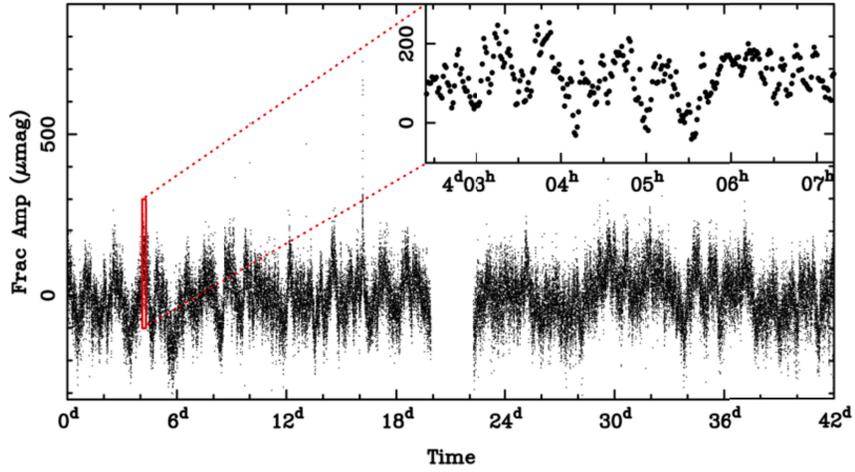}}
\caption{Processed 42-d portion of $\theta$\,Cyg light curve used for pulsation 
analysis} 
\end{figure}

\begin{figure}%[tb]
\center{\includegraphics[width=0.9\columnwidth]{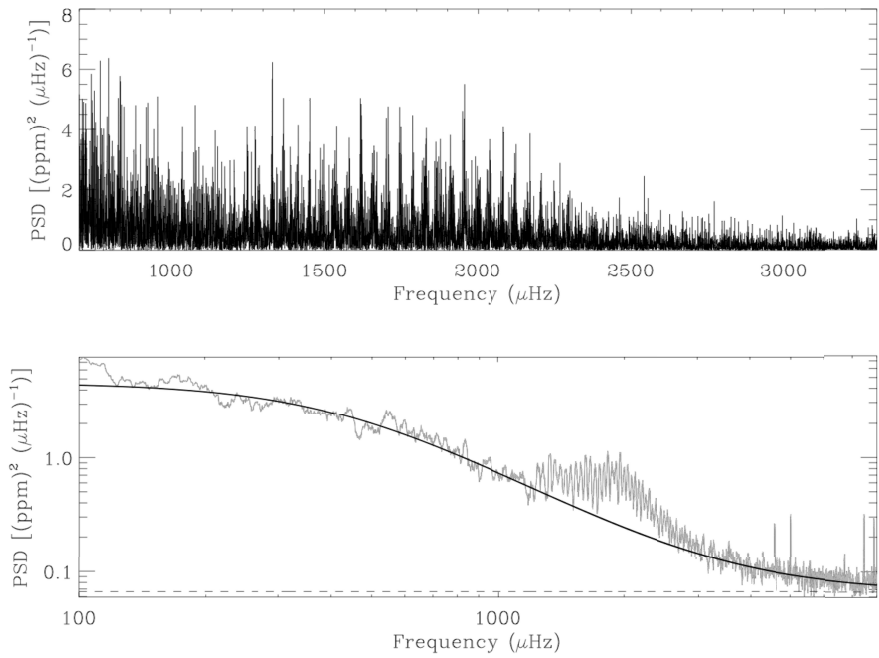}}
\caption{Fourier analysis of data showing power excess above background between 
1200 and 2500\,$\mu$Hz (bottom) and many solar-like oscillation peaks (top)} 
\end{figure}

\section{$\theta$\,Cyg models}

To determine whether $\gamma$\,Dor-like g modes could be present in a star such as  
$\theta$\,Cyg, we calculated stellar models that match the $L$ and $T_{\rm eff}$ 
inferred from the ground-based observations.  We calculated the evolutionary 
tracks of two models, one with mass 1.38\,M$_{\odot}$ and $Z=0.017$, close to the 
present photospheric $Z$ of the Sun (that has decreased from its initial value due 
to diffusive settling during the Sun's lifetime)\footnote{$Z$ is the mass fraction of elements heavier than hydrogen and helium} using the Grevesse \& Noels 
(1993) solar mixture, and a model with mass 1.29\,M$_{\odot}$ and $Z=0.013$, close 
to the Asplund et al. (2005) solar abundance determination.  We used the OPAL 
(Iglesias \& Rogers 1996) opacities, initial helium mass fraction $Y=0.28$, and 
mixing length/pressure scale height ratio 1.9.  We did not include diffusive 
settling or convective overshoot.   Fig.\,4 shows an HR~diagram of the evolution 
tracks and a box bounding the$ L-T_{\rm eff}$ observational constraints from the 
literature.  Table\,1 summarizes the observational constraints and the properties 
of two models marked along the evolution tracks that lie at the top and bottom of 
the luminosity bounds, which we chose to analyse for pulsations.  The radius of the 
1.38-M$_{\odot}$ model at the top of the box is a little too high, and the radius 
of the 1.29-M$_{\odot}$ model at the bottom of the box is a little too low, but we 
chose these models to bracket the extremes in convection zone depth, which is an 
important property for $\gamma$\,Dor pulsation driving (see discussion below), 
given the luminosity constraint. 

\begin{figure}%[tb]
\center{\includegraphics[width=0.7\columnwidth]{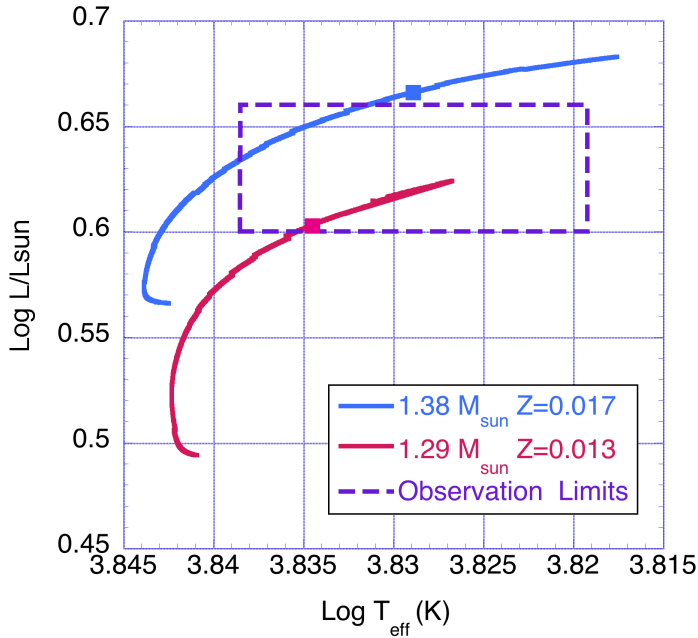}}
\caption{HR~diagram showing a box with observations and evolutionary tracks of 
1.38-M$_{\odot}$ and 1.29-M$_{\odot}$ models.  The square point on each track 
locates the two models analysed for $\gamma$\,Dor and solar-like oscillations} 
\end{figure}

The 1.38-M$_{\odot}$ higher-metallicity model has a deeper convective envelope, 
with temperature at the base of the envelope convection zone near 500\,000\,K, whereas 
the 1.29-M$_{\odot}$ model with 
lower metallicity has a convective envelope base temperature of $\sim$450\,000\,K.   
For $\gamma$\,Dor stars, the convective envelope base temperature that optimizes 
the growth rates and number of unstable g modes is predicted to be about 
300\,000\,K (see Guzik et al.\,2000 and Warner et al.\,2003).   Therefore, the 
majority of $\gamma$\,Dor g-mode pulsators are expected to 
have somewhat larger mass ($\sim$1.6\,M$_{\odot}$)  for solar metallicity and therefore shallower 
convective envelopes than $\theta$\,Cyg.   For models with convective envelopes 
that are too deep, the radiative damping below the convective envelope quenches 
the pulsation driving; for models with convective envelopes that are too shallow, 
the convective time scale becomes shorter than the g-mode pulsation periods, and 
convection can adapt during the pulsation cycle to transport radiation, making the 
convective blocking mechanism ineffective for driving the pulsations.

\subsection{Expectations for $\gamma$\,Dor g-mode Oscillations}

We analysed the models for pulsations using the Pesnell (1990) non-adiabatic
pulsation code, which also was used by Guzik et al.\,(2000) and Warner et al.\,(2003) to 
explain the pulsation driving mechanism for $\gamma$\,Dor pulsations and to predict 
the instability strip location.   The results are summarized in Table~2.  For the 
1.38-M$_{\odot}$ model with the deeper convective envelope, only one $l=1$  
high-order g mode is predicted to be pulsationally unstable.    For the
1.29-M$_{\odot}$ model with a shallower convective envelope, several  $l=1$ and 
$l=2$ modes with periods in the range $0.33 - 1$\,d are predicted to be 
unstable.  Therefore, if lower metallicities, appropriate for the Asplund et al. 
(2005) solar abundances, are adopted, $\theta$\,Cyg has potential to show 
$\gamma$\,Dor g-mode pulsations, as well as the already-observed solar-like 
oscillations. This star would then become the first known hybrid $\gamma$\,Dor/solar-like oscillator.  
Longer time series observations will be required to improve the signal-to-noise 
ratio and separate the g modes from the granulation signal at low frequency to 
detect the potential g modes.  On the other hand, if this star does not show g 
modes, such observations would support the higher metallicity more appropriate for
the older solar abundances, also favored by helioseismology (see, e.g., 
Basu \& Antia 2008; Guzik \& Mussack 2010), and/or help to constrain the location 
of the red edge of the $\gamma$\,Dor instability strip. Therefore, continued 
observations of $\theta$\,Cyg could shed light both on $\gamma$\,Dor g-mode 
instability, and on the solar abundance problem. 

\begin{table}
\caption{g-mode Predictions for $\theta$\,Cyg Models}
\begin{center}
\begin{tabular}{llll} 
\hline
 &   g-mode Properties	& 1.38-M$_{\odot}$  Model 1& 1.29-M$_{\odot}$  Model 2  
\\
\hline
\\
%$l$=1 & g-mode radial orders for P .5 to 3 d  $n$ & 15-60  &  15-50  \\
$l$=1  & period spacings (d)   &  $\sim$ 0.040    &  $\sim$ 0.038  \\
& radial orders  $n$ & 22  &  18 to 25  \\
& growth rates/period$^{a}$  &  4.5e-09   &   $<$1.0e-07   \\
& periods (d)   &  0.77    &  0.58 to 0.95  \\

\hline
\\
%$l$=2  & g-mode radial orders for  period 0.5 to 3 d  $n$ & 23-100  &  26-80 
 $l$=2  & period spacings (d)   &  $\sim$ 0.027  &  $\sim$ 0.025  \\
 & radial orders  $n$ & no unstable modes  & 20 to 28   \\
 & growth rates/period$^{a}$    &     &   $<$1.0e-7  \\
 & periods (d)   &     &  0.37 to 0.55  \\

\hline
\end{tabular}
\tablenotetext{a}{Calculated $\gamma$ Dor growth rates can be as high
as 1.0e-05/period}
\end{center}
\end{table}

\subsection{Large Separations of Solar-like Oscillations}

Solar-like oscillations with high radial order exhibit characteristic large 
frequency separations, $\Delta\nu$, between $l=1$ and $l=0$ modes of the same radial order, 
and small separations, $\delta \nu$, between $l=2$ and $l=0$ and between 
$l = 1$ and $l = 3$ modes of consecutive radial order (e.g., Aerts et al. 2010).  
An autocorrelation analysis of the frequency separations in the $\theta$\,Cyg 
solar-like oscillations shows a peak at multiples of 42\,$\mu$Hz, interpreted to 
be half the large separation, $\frac{1}{2}\Delta\nu$ (Fig.\,5).   By comparison, 
half of the large frequency separation for the Sun is 67.5\,$\mu$Hz.   Fig.\,6 
shows a histogram of the number of frequency spacings per one-$\mu$Hz  bin for all 
of the calculated p-mode frequencies between 1200 and 2500\,$\mu$Hz for our 1.38- 
and 1.29-M$_{\odot}$ models.  The 1.38-M$_{\odot}$  model has a mean 
$\frac{1}{2}\Delta\nu = 38.9 \pm 1.4$\,$\mu$Hz, while the 1.29-M$_{\odot}$ model 
has mean $\frac{1}{2}\Delta\nu = 43.2 \pm 1.2$\,$\mu$Hz; the results for the two models bracket 
the observed large separation.

\begin{figure}%[tb]
\center{\includegraphics[width=0.6\columnwidth]{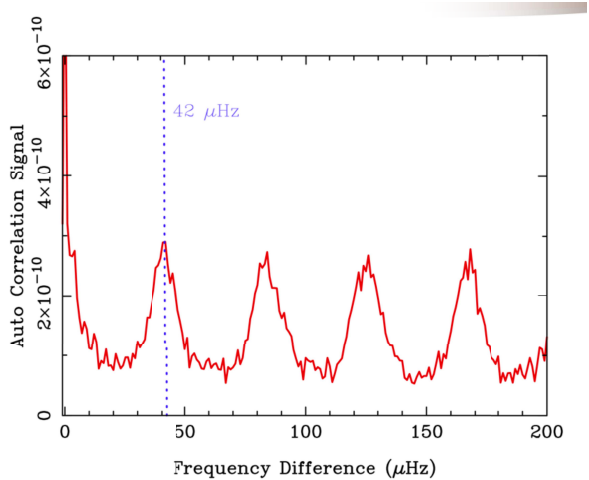}}
\caption{Autocorrelation of spacing of p-mode oscillations, showing a 42-$\mu$Hz 
peak interpreted as half of the large frequency separation between adjacent 
$l$=1 and $l$=0 modes} 
\end{figure}
 
\begin{figure}
\center{\includegraphics[width=0.6\columnwidth]{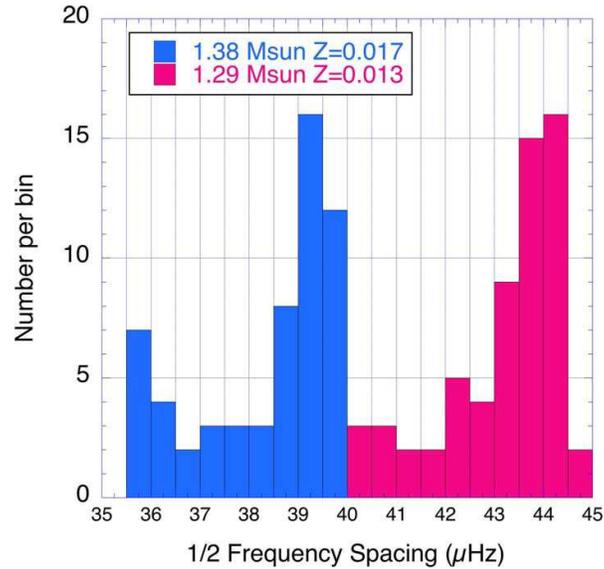}}
\caption{Separations between calculated p-mode oscillations for 1.38- and
1.29-M$_{\odot}$ models, binned into 1-$\mu$Hz intervals.  Half the large separation is 
$\sim$39 $\mu$Hz for the 1.38-M$_{\odot}$ model, and $\sim$44 $\mu$Hz for the
1.29-M$_{\odot}$ model, bracketing the observations at about 42 $\mu$Hz} 
\end{figure}

\section{Follow-up and Advantages of Optical Interferometry}

Follow-up {\it Kepler} photometry of $\theta$\,Cyg, as well as optical 
interferometry and spectroscopic abundance analyses, will be worthwhile.  The 
first data show solar-like p modes but no evidence (yet) of $\gamma$\,Dor gravity 
modes.  Our stellar models show that $\theta$\,Cyg is likely on, or very near, the 
red edge of the $\gamma$\,Dor instability strip.  An interferometric angular 
diameter combined with the Hipparcos parallax can narrow the constraints for 
$\theta$\,Cyg to better accuracy than spectroscopic/photometric analyses alone.  
Improved constraints are essential for g-mode predictions, since a lower radius 
from interferometry would drive model solutions to lower metallicity for a given 
luminosity, so that the convective envelope may be shallow enough to allow some 
unstable g-mode pulsations.  Resolving and tracking the 0.35\,M$_{\odot}$ M-dwarf 
companion for most of its very long orbital period, or finding one or more planets 
in closer orbits around $\theta$\,Cyg would help further to constrain its mass. 
$\theta$\,Cyg {\it Kepler} data for p modes and, if found, for g modes, combined with 
detailed abundance analyses, has high potential to shed light on the ongoing solar 
abundance problem.

\acknowledgements
The authors acknowledge the entire Kepler team for acquiring this data using the 
custom aperture.  G.H. acknowledges support from the Austrian FWF Project P21205-N16.  J.G. acknowledges support from the {\it Kepler} Guest Observer grant KEPLER08-0013, and thanks the conference organizers for the opportunity to present this work.

%\bibliography{aspauthor}

\bibliography{}

{}

\end{document}